\documentclass[12pt]{iopart}                                                   
\usepackage{psfig}                                                                 
\usepackage{epsfig}                                                                
\jl{1}                                                                             
\begin{document}
\title{Statistical properties of resonance widths for open Quantum Graphs}
\author{Tsampikos Kottos$^{1}${\footnote {corresponding author:
tsamp@chaos.gwdg.de}}
and Holger Schanz$^{2}$}
\address{$^{1}$Max-Planck-Institut f\"ur Str\"omungsforschung, G\"ottingen,
Germany}
\address{$^{2}$Institut f{\"u}r Nichtlineare Dynamik, Universit{\"a}t
G\"ottingen, Germany}
\date{\today }

\begin{abstract}
We connect quantum compact graphs with infinite leads, and turn them into scattering 
systems. We derive an exact expression for the scattering matrix, and explain how it 
is related to the spectrum of the corresponding closed graph. The resulting expressions
allow us to get a clear understanding of the phenomenon of resonance trapping due to
over-critical coupling with the leads. Finally, we analyze the statistical properties 
of the resonance widths and compare our results with the predictions of Random Matrix 
Theory. Deviations appearing due to the dynamical nature of the system are pointed out
and explained.  \\

\hspace {-0.4cm} submitted to Waves in Random Media -- special issue on Graphs
\end{abstract}
\maketitle


\section{\bf Introduction}
\label{sec:introduction}

Quantum graphs of one-dimensional wires connected at nodes were introduced already more 
than half a century ago to model physical systems. Depending on the envisaged application 
the precise formulation of the models can be quite diverse and ranges from solid-state 
applications to mathematical physics 
\cite{A85,FJK87,ML77,A81,CC88,NYO94,Ibook,MT01,Z98,V98,E95,R83}. Lately, quantum graphs 
attracted also the interest of the quantum chaos community because they can be viewed 
as typical and yet relatively simple examples for the large class of systems in which 
classically chaotic dynamics implies universal correlations in the semiclassical limit 
\cite{KS97,KS99,A99,BK99,SS00,T00,KS01,BSW02,BG00,K01}. Up to now we have only a limited 
understanding of the reasons for this universality, and quantum graph models provide a 
valuable opportunity for mathematically rigorous investigations of the phenomenon. In 
particular, for quantum graphs an exact trace formula exists \cite{R83,KS97,KS99} which 
is based on the periodic orbits of a mixing classical dynamical system. Moreover, it is 
possible to express two-point spectral correlation functions in terms of purely combinatorial 
problems \cite{BK99,SS00,T00,KS01,BSW02} and to use this in order to investigate the 
origin of the connection between Random Matrix Theory (RMT) and the underlying classical 
chaotic dynamics.

By attaching infinite leads at the vertices, we get non-compact graphs, for which a
scattering theory can be developed \cite{KS00,BG01,TM01}. The quantum scattering matrix 
for such systems can be written explicitly, and the corresponding resonances can be 
calculated easily. It is the purpose of this paper to review their statistical properties 
and analyze their dependence on the boundary conditions that we impose at the vertices 
\cite{FS96b,FS96,FS97}. At the same time we apply the reaction-matrix formalism and for
the first time we give a semiclassical interpretation of this theory in terms of the
classical trajectories of the graph.

The paper is structured in the following way. In section \ref{sec:definitions}, the 
mathematical model is introduced and the main definitions are given. Section 
\ref{sec:s-matrix} is devoted to the derivation of the scattering matrix for graphs. 
In section \ref{sec:rel} we discuss the relation between the S-matrix of an open graph 
and the spectrum of the corresponding closed system (reaction-matrix formalism). The 
resonance width distribution for various types of graphs are presented and discussed 
in section \ref{sec:reson} and deviations from RMT predictions \cite{FS96b,FS96,FS97} 
are analyzed. In the same section we also exhibit and analyze the phenomenon of resonance 
trapping occurring in specific graphs. Our conclusions are summarized in section 
\ref{sec:conclusions}.


\section{\bf Quantum Graphs: Definitions}
\label{sec:definitions}

We start by considering a {\it compact} graph ${\cal G}$. It consists of $V$ {\it vertices} 
connected by $B$ {\it bonds}. The number of bonds which emanate from each vertex $i$ defines 
the valency $v_i$ of the corresponding vertex (for simplicity we will allow only a single 
bond between any two vertices). The graph is called $v-regular$ if all the vertices have 
the same valency $v$. The total number of bonds is $B={1\over 2} \sum_{i=1} ^V v_{i}$. 
Associated to every graph is its connectivity matrix $C$. It is a square matrix of size 
$V$ whose matrix elements $C_{i,j}$ take the values $1$ if the vertices $i,j$ are connected 
with a bond, or $0$ otherwise. The bond connecting the vertices $i$ and $j$ is denoted by 
$b \equiv (i,j)$, and we use the convention that $i<j$. It will be sometimes convenient 
to use the ``time reversed" notation, where the first index is the larger, and $\hat b 
\equiv (j,i)$ with $j>i$. We shall also use the directed bonds representation, in which 
$b$ and $\hat b$ are distinguished as two directed bonds conjugated by time-reversal. We 
associate the natural metric to the bonds, so that $x_{i,j}\ (x_{j,i})$ measures the
distance from the vertex $i\ (j)$ along the bond. The length of the bonds are denoted 
by $L_{b}$ and unless stated otherwise we shall assume that they are {\it rationally 
independent}. The mean length is defined by $\left<L \right>\equiv (1/B) \sum_{b=1}^B L_b$ 
and in all numerical calculations below it will be taken to be $1$. In the {\it directed-
bond} notation $L_{b} = L_{\hat b}$. Finally, we define the total length of the graph 
${\cal L}=\sum_{b=1}^B L_b$. 

The {\it scattering} graph ${\tilde {\cal G}}$ is obtained by adding leads which extend 
from $M (\leq V)$ vertices to infinity. For simplicity we connect at most one lead to 
any vertex. The valency of these  vertices increases to ${\tilde v}_i =v_i+1$. The $M$ 
leads are denoted by the index $i$ of the vertex to which they are attached while $x_i$ 
now measures the distance from the vertex along the lead $i$.

The Schr\"odinger operator (with $\hbar=2m=1$) is defined on the graph ${\tilde {\cal G}}$ 
in the following way: On the bonds $b$, the components $\Psi_b$ of the total wave function 
$\Psi$ are solutions of the one - dimensional equation 
\begin{equation}
\label{schrodinger}
\left( -i{\frac{{\rm d\ \ }}{{\rm d}x}}-A_b\right) ^2\Psi _b(x)=k^2\Psi _b(x), \,\,\,
\,\,\,\,\,\,\,\,\, \bigskip\ b=(i,j)
\end{equation}
where $A_b$ (with $\Re e(A_{b})\ne 0$ and $A_{b}= -A_{\hat b}$) is a ``magnetic vector 
potential" which breaks the time reversal symmetry. In most applications we shall assume 
that all the $A_{b}$'s are equal and the bond index will be dropped. The components of 
the wave functions on the leads, $\Psi_i(x)$, are solutions of
\begin{equation}
\label{schrodinger1}
-{\frac{{\rm d^2\ \ }}{{\rm d}x^2}}\Psi _i(x)=k^2\Psi_i(x), \,\,\,\,\,\,\,\,\,\,\,\, 
\bigskip\ i=1,...,M.
\end{equation}
At the vertices, the wavefunction satisfies boundary conditions which ensure current
conservation. To implement the boundary conditions, the components of the wave function
on each of the bonds $b$ and the leads $i$ are expressed in terms of counter propagating
waves with a wave-vector $k$:
\begin {eqnarray}
&&{\rm On\ the \  bonds:\ } \Psi_{b} = a_{b} {\rm e}^{i(k+A_{b})x_{b}}
+c_{ b} {\rm e}^{i(-k+A_{b})x_{b}} \nonumber  \\
&&{\rm On\  the \ leads\ :\ } \Psi_i =  I_{i} {\rm e}^{-ikx_{i}} + O _{i}
{\rm e}^{ ikx_{i}} \ .
\label{wfun1}
\end{eqnarray}
The amplitudes $a_b,c_b$ on the bonds and $I_i,O_i$ on the lead are related  by
\begin{eqnarray}
\label{vertexcond}
{\hspace {-20mm}}
{\left ( \begin {array} {c}
O_i  \\  a_{i,j_1}\\ \cdot  \\   a_{i,j_{v_i}}
\end {array}
\right )} =
\Sigma^{(i)}
{\left ( \begin{array} {c}
I_i  \\ c_{j_1,i} \\ \cdot  \\ c_{j_{v_i},i}
\end {array} \right ) }, \quad\quad \quad\quad
\Sigma^{(i)}=
{\left (
\begin{array}{cccc}
\rho^{(i)}  & \tau^{(i)}_{j_1} & \cdot   & \tau^{(i)}_{j_{v_i}} \\
\tau^{(i)}_{j_1} & \tilde \sigma^{(i)}_{j_1,j_1} & \cdot  &
\tilde \sigma^{(i)}_{j_1,j_{v_i}} \\
\cdot  & \cdot &\cdot &\cdot \\
\tau^{(i)}_{j_{v_i}} & \tilde \sigma^{(i)}_{{j_{v_i}},j_1} &
\cdot & \tilde \sigma^{(i)}_{j_{v_i},j_{v_i}} \\
\end{array}
\right)}.
\end{eqnarray}
These equalities impose the  boundary conditions at the vertices. The vertex scattering
matrices $\Sigma^{(i)}_{j,j'}$, are $\tilde v_i \times \tilde v_i$ unitary symmetric
matrices, and $j,j'$ go over all the $v_i$ bonds and the lead which emanate from $i$.
The unitarity of $\Sigma^{(i)}$ guarantees current conservation at each vertex.

On the right hand side of (\ref{vertexcond}), the vertex scattering matrix $\Sigma^{(i)}$
was written explicitly in terms of the vertex reflection amplitude $\rho^{(i)}$, the
lead-bond transmission amplitudes $\{ \tau^{(i)}_{j}\}$, and the $v_i\times v_i$ bond-bond
transition matrix $\tilde \sigma^{(i)}_{j,j'}$, which is {\it sub unitary} ($|\det \tilde
\sigma^{(i)}|<1 $), due to the coupling to the leads. Vertices which are not coupled
to leads have $\rho^{(i)}=1 ,\ \tau^{(i)}_j =0$, while the bond-bond transition matrix
$\tilde \sigma^{(i)}_{j,j'}$ is unitary.

Graphs for which there are no further requirements on the $\Sigma^{(i)}$ shall be referred 
to as {\it generic}. In this case the vertex scattering matrices $\Sigma^{(i)}$ are chosen 
from the ensemble of unitary symmetric $\tilde v_i \times \tilde v_i$ random matrices, as 
explained in \cite{KS01}. However, often it is more convenient to compute the vertex scattering
matrices from the requirement that the wave function is continuous and satisfies current
conservation at all vertices \cite{KS97,KS99}. In this case the resulting $\tilde \Sigma^{(i)}$ 
matrices read \cite{KS00}:
\begin {equation}
\label{Neumann}
\tilde \sigma^{(i)} _{j,j'}= {(1+e^{-i\omega _i})\over {\tilde v}}  -\delta
_{j,j'};
\quad \tau^{(i)}_j =   { (1+e^{-i\omega _i}) \over {\tilde v}};\quad
\rho^{(i)} ={(1+e^{-i\omega _i}) \over {\tilde v}}-1\
\end{equation}
where $\omega_i=2\arctan \frac{\lambda_i}{{\tilde v_i}k}$. The parameters $0\le\lambda_i
\le\infty$ are free parameters which determine the boundary conditions. We shall refer 
to the $\lambda_i$ as the {\it vertex scattering potential}. The special case of zero 
$\lambda_i$ corresponds to Neumann boundary conditions. Dirichlet boundary conditions 
result from $\lambda_i = \infty$. A finite value of $\lambda_i$ introduces a new length 
scale. It is natural therefore, to interpret it in physical terms as a representation of 
a local impurity or an external fields \cite{ES89}. We finally note that the above model 
can be considered as a generalization of the Kronig-Penney model to a multiply connected, 
yet one dimensional manifold.


\section{\bf The S-matrix for Quantum Graphs}
\label{sec:s-matrix}

It is convenient to discuss first graphs with leads connected to all the vertices $M=V$.
The generalization to an arbitrarily number $M\leq V$ of leads (channels) is straightforward 
and will be presented at the end of this section.

To derive the scattering matrix, we first write the  bond wave functions using the two 
representations which are conjugated by ``time reversal":
\begin{eqnarray}
\label{wfun2}
\Psi_{b}(x_b) &=& a_{b} {\rm e}^{i(k+A_{b})x_{b}}
+c_{ b} {\rm e}^{i(-k+A_{b})x_{b}}\  = \nonumber \\
\Psi _{\hat b}(x_{\hat b}) &=&a_{\hat b}{\rm e}^{i(k+A_{\hat b})x_{\hat b}}+
c_{\hat b}{\rm e}^  {i(-k+A_{\hat b})x_{\hat b}}\  = \nonumber  \\
&=&a_{\hat b}{\rm e}^  {i(-k-A_{\hat b})x_b}{\rm e}^  {i(k+A_{\hat b})L_b} +
c_{\hat b}{\rm e}^  {i(-k+A_{\hat b})L_b}{\rm e}^  {i(k-A_{\hat b})x_b} \ .
\end{eqnarray}
Hence,
\begin{equation}
c_b=a_{\hat b}{\rm e}^  {i(k+A_{\hat b})L_b},\,\,\,\,
a_b=c_{\hat b}{\rm e}^  {i (-k+A_{\hat b})L_b }.\label{wfun3}
\end{equation}
In other words, but for a phase factor, the outgoing wave from the vertex $i$ in the
direction $j$ is identical to the incoming wave at $j$ coming from $i$.

Substituting $a_b$ from Eq.~(\ref{wfun3}) in Eq.~(\ref{vertexcond}), and solving for
$c_{i,j}$ we get
\begin{eqnarray}
\label{wfun4}
c_{i,j'}&=&\sum_{r,s} \left({\bf 1}-\tilde S_B(k;A) \right)^{-1}_{(i,r),(s,j)} D_{(s,j)}
\tau _s^{(j)} I_j \nonumber \\
O_i&=&\rho^{(i)} I_i + \sum_{j'}\tau_{j'}^{(i)} c_{ij'}
\end{eqnarray}
where ${\bf 1}$ is the $2B\times 2B$ unit matrix. Here, the ``bond scattering matrix''
${\tilde S}_b$ is a sub-unitary matrix in the $2B$ dimensional space of directed bonds
which propagates the wavefunctions. It is defined as $\tilde S_B(k,A)=D(k;A)\tilde R $,
with
\begin{eqnarray}
\label{DandT}
D_{ij,i^{\prime }j^{\prime }}(k,A) &=&\delta _{i,i^{\prime }}\delta_{j,j^{\prime }}
{\rm e}^{ikL_{ij}+iA_{i,j}L_{ij}} \\
\ \tilde R_{ji,nm} &=&\delta _{n,i}C_{j,i}C_{i,m}{\tilde \sigma}_{ji,im}^{(i)}.
\nonumber
\end{eqnarray}
$D(k,A)$ is a  diagonal unitary matrix which depends only on the metric properties of the 
graph, and provides a phase which is due to free propagation on the bonds. The sub-unitary 
matrix $\tilde R$ depends on the connectivity and on the bond-bond transition matrices 
${\tilde \sigma}$. It  assigns a scattering amplitude for transitions between connected 
directed bonds. $\tilde R$ is sub-unitary, since
\begin{equation}
\label{subR}
|\det \tilde R| = \prod_{i=1}^V |\det \tilde \sigma ^{(i)}|<1.
\end{equation}

Replacing $c_{i,j'}$ in the second of Eqs.~(\ref{wfun4}) we get the following relation
between the outgoing and incoming amplitudes $O_i$ and $I_j$ on the leads:
\begin{equation}
\label{inout}
O_i = \rho^{(i)} I_i + \sum_{j'j r s}\tau_{j'}^{(i)}
\left({\bf 1}-\tilde S_B(k;A) \right )^{-1}_{(i,r),(s,j)} D_{(s,j)} \tau_s^{(j)} I_j\,.
\end{equation}
Combining (\ref{inout}) for all leads $i=1,\dots ,V$, we obtain the unitary $V\times V$
scattering matrix $S^{(V)}$,
\begin{equation}
\label{scatmat}
S^{(V)}_{i,j}  = \delta_{i,j} \rho^{(i)} + \sum_{r,s} \tau^{(i)}_r \left ({\bf 1}-\tilde 
S_B(k;A) \right )^{-1}_{(i,r),(s,j)} D_{(s,j)} \tau _s^{(j)}.
\end{equation}
Finally, a multiple scattering expansion for the $S-$matrix is obtained by substituting
\begin{equation}
\label{multi}
({\bf 1}-\tilde S_B(k;A))^{-1}= \sum_{n=0}^{\infty} \tilde S_B ^n(k;A)
 \,\,\,\,\,\,
 \end{equation}
into Eq.~(\ref{scatmat}). The expansion (\ref{multi}) converges absolutely on the real
$k-$axis since for generic graphs the spectrum of ${\tilde S}$ is strictly inside the
unit circle (or else, for $k + i \epsilon$, with $\epsilon >0$ arbitrarily small).

The scattering matrix can be decomposed into two parts $S^{(V)}(k) =\langle S^{(V)}
\rangle_k +S^{fl}(k)$ which are associated with two well separated time scales of the
scattering process. $\langle S^{(V)} \rangle_k$ can be evaluated by averaging term by
term the $S-$matrix resulting from the substitution of the multiple scattering expansion
(\ref{multi}) in (\ref{scatmat}). Since all terms, but for the prompt reflections, are 
oscillatory functions of $k$, they average to zero (we assume that the $k-$window over 
which the averaging is performed is sufficiently large i.e. larger than $2\pi/L_{\rm min}$) 
and thus we get that $\langle S^{(V)} \rangle_k=S^D= \delta_{i,j} \rho^{(i)}$. The 
fluctuating component of the $S$ matrix, $S^{fl}(k)$, starts by a transmission from the 
incoming lead $i$ to the bonds $(i,r)$ with transmission amplitudes $\tau^{(i)}_{r}$. 
The wave gains a phase ${\rm e}^{i(k+A_{b})L_{b}}$ for each bond it traverses and a 
scattering amplitude $\tilde \sigma^{(i)}_{r,s}$ at each vertex. Eventually the wave 
is transmitted from the bond $(s,j)$ to the lead $j$ with an amplitude $\tau _s^{(j)}$.
Explicitly,
\begin{equation}
S^{(V)}_{i,j}  = \delta_{i,j} \rho^{(i)} +  \sum_{t \in {\cal
T}_{i\rightarrow j}}
{\cal B}_{t} {\rm e}^{i (k l_t +  \Theta_t)}
\label {sexplicit}
\end{equation}
where ${\cal T}_{i\rightarrow j}$ is the set of the trajectories on $\tilde
{\cal G}$ which lead from $i$ to $j$.  ${\cal B}_{t}$ is the amplitude
corresponding to a path $t$ and is given by the product of all
vertex-scattering matrix elements encountered along $t$.
Length and directed length of the path are $l_t= \sum_{b\in t}L_b$ and
$\Theta_t= \sum_{b\in t} L_bA_b$, respectively. Thus the scattering amplitude
$S^{(V)}_{i,j}$ is a sum of a large number of partial amplitudes, whose
complex interference brings about the typical irregular fluctuations of
$|S^{(V)}_{i,j}|^2$ as a function of $k$.

We finally comment that the formalism above can be easily modified for graphs where
not all the vertices are attached to leads. If the vertex $l$ is not attached, one
has to set $\rho^{(l)}=1, \tau^{(l)}_j =0$  in the definition of $\Sigma^{(l)}$.
The dimension of the scattering matrix is then changed accordingly.


\section{\bf Relations between spectrum and S-matrix }
\label{sec:rel}

\subsection{\bf Secular equation}

There exists an intimate link between the scattering matrix and the spectrum of the
corresponding closed graph. It manifests the exterior-interior duality for graphs
\cite{KS97,KS00}. The spectrum of the closed graph is the set of wave-numbers for which
$S^{(V)}$ has $+1$ as an eigenvalue. This corresponds to a solution where no currents
flow in the leads so that the conservation of current is  satisfied on the internal
bonds. $1$ is in the spectrum of $S^{(V)}$ if
\begin{equation}
\label{phsca4}
\zeta_{\cal G}(k)=\det\left[{\bf 1} -S^{(V)}(k)\right] = 0 \quad .
\end{equation}
Eq.~(\ref{phsca4}) can be transformed in an alternative  form
\begin{equation}
\label{phscaa} {\hspace {-15mm}}
\zeta_{\cal G}(k)= \det [ {\bf 1} -\rho ]
{\det[{\bf 1}-D(k)R] \over \det[{\bf 1}-D(k) {\tilde R}]} = 0 \quad ; \quad
R_{i,r;s,j} = \tilde R_{i,r;s,j} +\delta_{r,s}
{\tau^{(r)}_{i}\tau^{(r)}_{j}\over 1-\rho^{(r)} }.
\end{equation}
which is satisfied once
\begin{equation}
\label{phscaa1}
\det[{\bf 1}-D(k)R] = 0 .
\end{equation}
In contrast to ${\tilde R}$, $R$ is a unitary matrix in the space of directed bonds,
and therefore the spectrum is real. Eq.~(\ref {phscaa1}) is the secular equation for 
the spectrum of the compact part of the graph \cite{KS97,KS99,KS00}.

\subsection{\bf Reaction-matrix formalism}\label{rmat}

Eq.~(\ref{phsca4}) allows to compute the spectrum of the closed graph from the $S-$matrix 
of the open graph. This relation can be inverted: it is possible to express the $S-$matrix 
at an arbitrary wavenumber $k$ in terms of all eigenvalues and eigenfunctions of the 
corresponding closed graph. This is achieved by applying to quantum graphs
methods which were developed in the context of nuclear physics: Wigner and
Eisenbud \cite{WE47} introduced in the late forties the reaction matrix as a
connecting element between a closed and the corresponding scattering system. The
projection-operator formalism of Feshbach \cite{Feshbach} allows a systematic
separation of the Hilbert space of a closed system from the embedding
scattering system and on this basis Weidenm{\"u}ller and coworkers developed a
theory expressing the S-matrix in terms of an effective non-Hermitian
Hamiltonian \cite{MW69} which contains the dynamics of the closed system and
the coupling to the scattering channels. This effective Hamiltonian is the
basis for the RMT results on resonance statistics that we will discuss in the
following sections, and therefore it is important to understand how
reaction-matrix theory and Weidenm{\"u}ller approach work in quantum graphs.

\begin{figure}
\center{
\psfig{figure=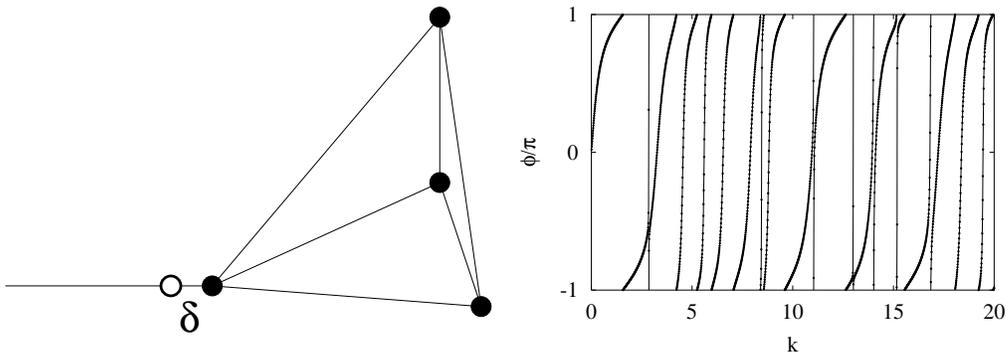,scale=0.55,angle=0}
\caption {\label{fig:th}
A tetrahedron with a single attached scattering channel (left) is used to
illustrate the reaction-matrix theory for graphs. The empty circle shows an
auxiliary vertex with Neumann b.~c. which is used for the derivation of
Eq.~\protect\ref{SK1}. On the right we compare the exact phase of the
S-matrix of this graph (full line) to an approximation computed with the
help of Eqs.~(\protect\ref{GF}), (\protect\ref{SK2}) from the lowest 500
eigenstates of the graph (dots).  } }
\end{figure}

In order to achieve this understanding we consider first the Green's function $G$
of a {\em closed} graph. It can be defined by the infinite spectral sum
\begin{equation}
\label{GF}
G(x_b,x_{b'};k)= \sum_{n=1}^{\infty}{\Phi_{n}(x_{b})\Phi_{n}^{*}(x_{b'})\over
k^{2}-k_{n}^{2}}\,,
\end{equation}
where $k_n$ denote the $n$th eigenvalue and $\Phi_{n}(x_b)$ the corresponding
eigenfunction at distance $x_b$ at the bond $b$. The graph will later be opened 
by attaching leads to the vertices $1\le j\le M$ on 
which we assume Neumann boundary conditions (the b.~c. at other vertices are 
arbitrary). We denote by $G_{jj'}$ the values of the Green's function on these 
vertices and introduce the Hermitian {\em reaction matrix}
\begin{equation}\label{KG}
K_{jj'}=k\,G_{jj'}\nonumber\,.
\end{equation}
We will show that $K$  is related to the S-matrix via
\begin{eqnarray}\label{SK1}
K&=&-\i(I-S)^{-1}(I+S)\,,
\end{eqnarray}
which is equivalent to
\begin{eqnarray}\label{SK2}
S&=&-(I+\i\,K)^{-1}(I-\i\,K)\,.
\end{eqnarray}
For technical reasons it is easier to derive Eqs.~(\ref{SK1}), (\ref{SK2}) in
a situation where the leads are attached to vertices $j$ with valency
$v_{j}=1$ and Neumann b.~c. An example is shown in Fig.~\ref{fig:th}.  As long
as we consider Neumann b.~c. this is no restriction of generality since the
length $\delta$ of the bond reaching a "dead end" at vertex $j$ in the closed
graph can be arbitrarily small, and it can easily be shown that in the limit
$\delta\to 0$ spectrum and eigenfunctions of the graphs with and without the
attached arm coincide.  Hence, after Eq.~(\ref{SK1}) is established with the
help of auxiliary vertices, these vertices can be disregarded. Note that due
to this construction $\rho^{(j)}$ in Eqs.~(\ref{scatmat}), (\ref{sexplicit})
is zero: in the scattering system the vertices to which the leads are formally
attached have perfect transmission and are effectively absent.

There are two independent ways to arrive at Eqs.~(\ref{SK1}), (\ref{SK2}).
The first simply repeats the corresponding arguments for billiard systems
\cite{Dit00,PSS01}.  In fact the case of a quantum graph is completely
equivalent up to the additional simplification that the attached leads do not
allow for multiple transversal modes.  However, here we will follow a
different route which is more specific to quantum graphs as it relies on the
exact representation of the corresponding S-matrix as a sum over classical
trajectories (see Eq.~(\ref{sexplicit})). There is a similar expansion of the
Green's function in terms of classical trajectories
\begin{equation}
\label{sclGF}
G(x,x';k)={1\over 2i k}\sum_{t:x'\to x}{\cal
B}_{t}\,\e^{i(kl_{t}+\Theta_{t})}\,,
\end{equation}
where $t$ denotes a trajectory leading from $x'$ to $x$ while ${\cal B}_{t}$,
$k\,l_{t}$, and $\Theta_{t}$ are defined as in Eq.~(\ref{sexplicit}). Skipping a
more mathematical derivation of Eq.~(\ref{sclGF}) we only point the analogy with 
the well known Green's function of a free particle $G_0$ in one-dimension
$G_{0}=(2ik)^{-1}\e^{i|x-x'|}$.  Eq.~(\ref{sclGF}) simply generalizes this
expression by adding the additional paths from $x'$ to $x$ which are due to
the multiply connected nature of the graph.

Let us now consider the specific matrix elements $G_{jj'}$ of the Green's function
 Eq.~(\ref{sclGF}), which appear in the definition of the reaction matrix Eq.~(\ref{KG}). 
It is our goal to represent them in terms of the scattering trajectories contributing 
to the S-matrix element $S_{jj'}$ according to Eq.~(\ref{sexplicit}). To this end
we note that the contribution of each trajectory is the same in the two expressions 
up to an overall factor $(2ik)^{-1}$ appearing in the Green's function. However, the 
set of contributing trajectories is different: in the open system we have $\rho^{(j)}
=0$ meaning that no backscattered trajectories from the auxiliary vertices $1\le j\le M$
contribute, while in the closed graph such trajectories do exist. In order 
to get their contributions also from Eq.~(\ref{sexplicit}) we consider multiple 
scattering at the graph with scattering trajectories reinjected an arbitrary number 
of times. Then it becomes indeed possible to represent all trajectories inside the 
closed graph as combinations of scattering trajectories and we find
\begin{equation}\label{GS}
G_{jj'}=
(2ik)^{-1}\left(2\delta_{jj'}
+4\sum_{n=1}^{\infty} (S^{n})_{jj'}\right)\,,
\end{equation}
where $n$ counts the multiple scattering at the graph. The first term ($n=0$)
corresponds to the zero-length trajectories from $j$ to $j$. Note
that $G_{jj'}$ actually denotes the limit of $G(x,x';k)$ as
$x\to j$ and $x'\to j'$ and that there are always two options to
reach a point in the vicinity of a vertex $j$: with or without preceeding
scattering at this vertex. Combining these options for $j$ and
$j'$ gives the prefactor $4$ ($2$ for $n=0$) in
Eq.~(\ref{GS}). After summing the geometric series in Eq.~(\ref{GS}) we arrive
at Eq.~(\ref{SK1}).

Not only have we established in this way an analogue of the well known
reaction-matrix formalism for quantum graphs, we have also given the first
semiclassical interpretation of this theory: our derivation of
Eq.~(\ref{SK1}) was entirely based on comparing the contributions of classical
trajectories summing to an S-matrix element or Green's function, respectively.

In Fig.~\ref{fig:th} we demonstrate numerically, how Eq.~(\ref{SK2}) can be
used to compute the S-matrix of a scattering graph from its corresponding closed
analogue. For the tetrahedron graph (see figure) we computed the lowest $500$ 
eigenvalues and eigenfunctions and using Eqs.~(\ref{GF}), (\ref{SK2}) we obtain 
an approximation for the $S$-matrix of the graph (see circles). For simplicity we
consider the case $M=1$ where the $S-$matrix reduces to a single phase. The
resulting data are in excellent agreement with the exact result (solid line)
calculated from Eq.~(\ref{scatmat}). We verified that the comparison is equally
successful for multiple leads and other types of graphs.

We continue by deriving the effective Hamiltonian for a quantum graph. To this
end we rewrite Eq.~(\ref{KG}) in the standard form
\begin{equation}\label{}
K_{jj'}=
\pi \sum_{n=1}^{\infty}W_{n,j}^*{I\over k^2-k_{n}^2}W_{n,j'}\,,
\end{equation}
where
\begin{equation}\label{W}
W_{n,j}(k)=\sqrt{k/\pi}\psi_{n,j}
\end{equation}
denotes the coupling of the internal level $n$ to the scattering lead
$j$. We see from Eq.~(\ref{W}) that these coupling constants are
proportional to the internal eigenfunction at the point where the lead is
attached, as it should be expected. Following \cite{FS97} we can now use some
algebra to rewrite Eq.~(\ref{SK2}) in the form
\begin{equation}\label{}
S=I-2\pi i\,W^{\dag}\,{I\over E-{\cal H}}\,W\,,
\end{equation}
where the S-matrix is expressed in terms of an effective Hamiltonian $\cal H$,
which has in the eigenbasis of the graph the matrix elements
\begin{equation}\label{heff}
{\cal H}_{nn'}=k_{n}^{2}\delta_{nn'}-i\pi (WW^{\dag})_{nn'}\,.
\end{equation}
Note that this Hamiltonian depends on energy via Eq.~(\ref{W}). Whenever this
dependence can be neglected, Eq.~(\ref{heff}) can be diagonalized to find the
resonance poles. In this regime, the predictions of RMT for the statistical 
properties of the resonances \cite{FS96b,FS96,FS97} are applicable. 


\section{\bf Statistical properties of resonance widths }
\label{sec:reson}

One of the basic concepts in the quantum theory of scattering are the resonances.
They represent long-lived intermediate states to which bound states of a closed
system are converted due to coupling to continua. On a formal level, resonances
show up as poles of the scattering matrix $S^{(M)}$ occurring at complex wave-numbers
$\kappa_n = k_n - \frac{i}{2}\Gamma_n$, where $k_n$ and $\Gamma_n$ are the position
and the width of the resonances, respectively. From (\ref{scatmat}) it follows that
the resonances are the complex zeros of
\begin{equation}
\label {resonancecond}
\zeta_{\tilde {\cal G}}(\kappa) = \det \left({\bf 1} -\tilde S_B(\kappa ;A)\right)=0 \ .
\end{equation}
The eigenvalues of $\tilde S_B=D(k;A)\tilde R$ are in the unit circle, and therefore
the resonances appear in the lower half of the complex $\kappa$ plane.

One important parameter that is associated with the statistical properties of resonance
widths is the Ericson parameter defined through the scaled mean resonance width as:
\begin{equation}
\label{eripar}
\left<\gamma\right>_k \equiv \left<{\Gamma \over \Delta}\right>_k
\end{equation}
where $\langle \cdot \rangle _k$ denotes spectral averaging and $\Delta$ is the mean 
spacing between resonances. The Ericson parameter defines whether the resonances are 
overlapping $(\langle \gamma\rangle_k >1)$ or isolated $(\langle \gamma\rangle_k <1)$.

\begin{figure}
\center{
\psfig{figure=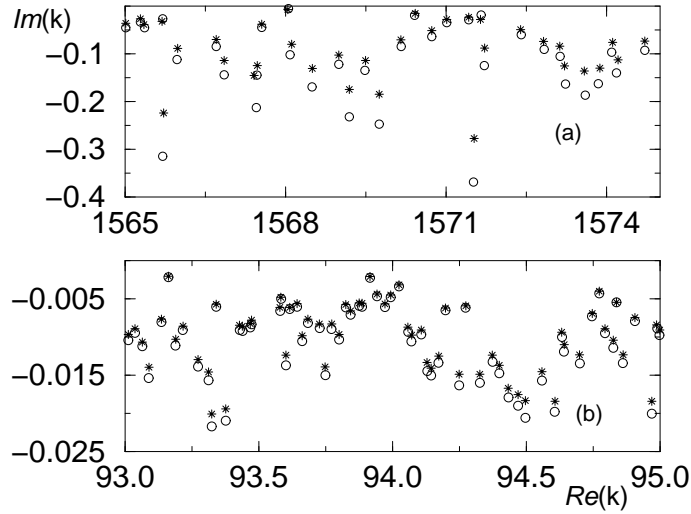,scale=0.5,angle=0}
\vspace*{5mm}
\caption {\label{fig1} Poles of the $S^{(V)}$-matrix for regular Neumann graphs.
The exact evaluated poles are indicated with $(\circ )$ while $(\star )$ are the results
of the perturbation theory (\ref{pertu3}): (a) $V=5$ and $v=4$ and (b) $V=15$ and $v=14$.}
}
\end{figure}

For relatively small values of the Ericson parameter a perturbation theory can be
carried out in order to evaluate the resonance widths. Indeed, the difference $\delta
R = R - \tilde R$ gets smaller as larger graphs are considered (for graphs with Neumann
boundary conditions it is easy to see that the difference is of order $\frac{1}{v}$).
That is, the leads are weakly coupled to the compact part of the graph. To lowest order, 
$(\delta R =0)$, the resonances coincide with the spectrum of the compact graph. Let 
$k_n$ be in the spectrum. Hence, there exists a vector $|n \rangle$ which satisfies 
the equation
\begin{equation}
D(k_n) R\  |n \rangle = 1\ |n \rangle \ .
\end{equation}
To first order in $\delta R$, the resonances acquire a width
\begin{equation}
\label{pertu3}
\delta \kappa_n = -i  {\langle n|D(k_n)\delta R|n\rangle \over \langle
n|L|n\rangle} \,.
\end{equation}
The above equation shows clearly that the formation of resonances is closely related
to the internal dynamics inside the scattering region which is governed by $\tilde S_B$.

To check the usefulness of Eq.~(\ref{pertu3}), we searched numerically for the true poles
for a few scattering graphs and compared them with the approximation~(\ref{pertu3}). In 
Fig.~\ref{fig1} we show the comparison for fully connected Neumann graphs with $V=5,15$
and $A=0$. As expected the agreement between the exact poles and the perturbative results 
improves as $v$ increases.

\subsection{\bf Resonance width distribution}

An important feature of the distribution of the resonances in the complex plane can be 
deduced by studying the secular function $\zeta_{\tilde {\cal G}}(\kappa)$. Consider 
$\zeta_{\tilde {\cal G}}(\kappa=0)$. If one of the eigenvalues of the matrix $\tilde R$ 
(\ref {DandT}) takes the value $1$, $\zeta_{\tilde {\cal G}}(k=0) =0$ and because of the 
quasi-periodicity of $\zeta_{\tilde {\cal G}}$, its zeros reach any vicinity of the real 
axis infinitely many times.  For $v$-regular Neumann graphs without magnetic vector
potential ($A=0$) the matrix $\tilde R$ has indeed an eigenvalue $1$, and therefore the 
distribution of resonance widths is finite in the vicinity of $\gamma=0$ for these systems. 
For generic graphs, the spectrum of $\tilde R$ is typically inside a circle of radius 
$\mu_{max} <1$. This implies that the poles are excluded from a strip just under the real 
axis, whose width can be estimated by
\begin{equation}
\label{gapKS}
\Gamma_{gap}=-2 \ln( |\mu_{max}|)/L_{max}\,,
\end{equation}
where $L_{max}$ is the  maximum bond length. The existence of a gap is an important 
feature of the resonance width distribution ${\cal P}(\gamma)$ for chaotic scattering
systems.

A similar argument was used recently in \cite{BG01} in order to obtain an {\it upper}
bound for the resonance widths. It is
\begin{equation}
\label{upl}
\Gamma_{max}\sim -2 \ln( |\mu_{min}|)/L_{min}\ ,
\end{equation}
where $\mu_{min}$ and $L_{min}$ are the minimum eigenvalue and bond length, respectively.

\begin{figure}
\begin{center}
\psfig{figure=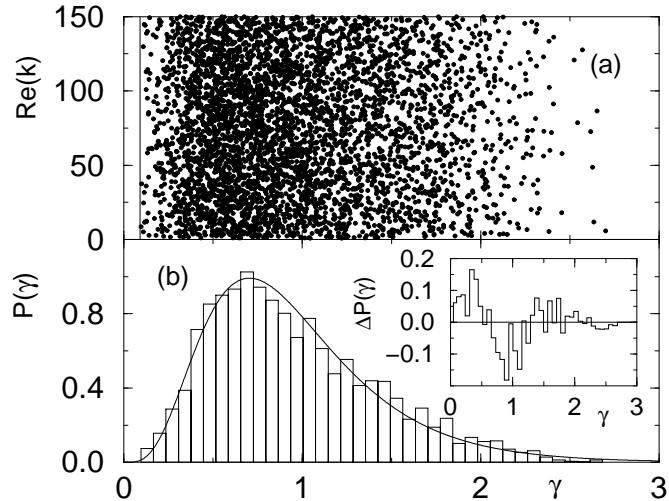,scale=0.5,angle=0}
\vspace*{5mm}
\caption{\label{fig2} (a) The 5000 resonances of a single realization of a complete $V=5$ 
graph with $A\ne 0$ and $M=V$. The solid line marks the position of the gap $\gamma_{\it gap}$.
(b) The distribution of resonance widths ${\cal P}(\gamma)$. The solid line is the RMT
prediction (\ref{FSpole}). The difference ${\cal P}(\gamma)-{\cal P}_{CUE} (\gamma)$ is 
shown in the inset.}
\end{center}
\end{figure}

The distribution of the complex poles  for a generic fully connected graph with $V=5$
is shown in Fig~\ref{fig2}a. The vertical line which marks the region from which resonances
are excluded was computed using (\ref {gapKS}).

Random matrix theory can provide a general expression for the distribution of
resonances.  Specifically, Fyodorov and Sommers \cite{FS96,FS97} proved that
the distribution of scaled resonance widths for the unitary random matrix
ensemble, is given by
\begin{equation}
\label{FSpole}
{\cal P}(\gamma) = \frac {(-1)^M}{\Gamma(M)} \gamma^{M-1} {d^M\over
d\gamma^M}
\left({\rm e}^  {-\gamma \pi g} {\sinh(\gamma\pi)\over(\gamma\pi)}\right)\,,
\end{equation}
where the parameter $g={2\over (1-\langle S^D\rangle_k^2)} -1$ controls the degree of
coupling with the channels (and it is assumed that $g_i=g \,\, \forall i=1,...M$).

In the limit of $M\gg 1 $, Eq.~(\ref{FSpole}) reduces to the following expression
\cite{FS97}
\begin{equation}
{\cal P}(\gamma)=
\left\{
\begin{array}{lll}
{M\over 2\pi \gamma^2}&\ {\rm for}\ &{M\over \pi(g+1)}<
\gamma<{M\over\pi(g-1)} \\\nonumber
\ 0&\ \ \ \ & \ \ \ \ \ \  {\rm otherwise} \nonumber
\end{array}
\right.  \ .
\label{largeM}
\end{equation}
It shows that in the limit of large number of channels there exists a strip in the
complex $\kappa-$ plane which is free of resonances. This is in agreement with
previous findings \cite{GR89,M75,SZ88}. In the case of maximal coupling i.e. $g=1$,
the power law (\ref{largeM}) extends to  infinity, leading to divergencies of the
various moments of $\gamma$'s. Using  (\ref{FSpole}) we recover the well known
Moldauer-Simonius relation \cite{M75} for the mean resonance width \cite{FS97}
\begin{equation}
\label{SMrw}
\langle \gamma\rangle _k = -\frac {\sum_{i=1}^{V} \ln (|\langle S^{D}\rangle_k|^2)}
{2\pi}.
\end{equation}

\begin{figure}
\begin{center}
\psfig{figure=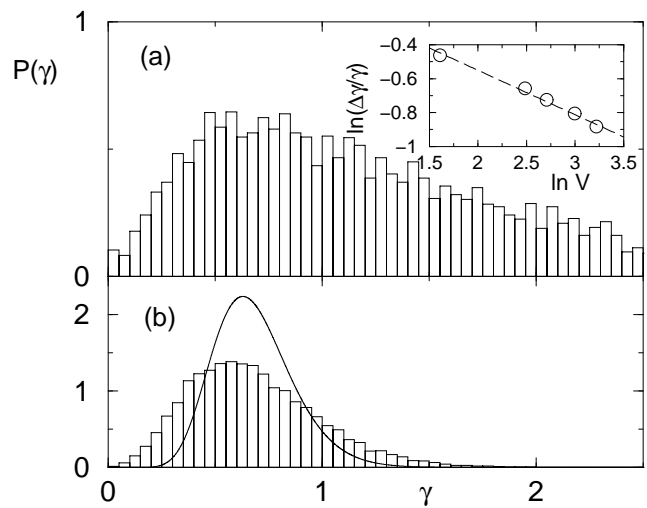,scale=0.5,angle=0}
\vspace*{5mm}
\caption {\label{fig3}  (a) Resonance width distribution ${\cal P}(\gamma)$ for fully 
connected Neumann graphs with (a) $A=0$ and $M=V=5$; (b) $A\neq 0$ and $M=V=15$. The 
solid line is the RMT prediction (\ref{FSpole}). In the inset we report the relative 
ratio $\Delta \gamma \over \left<\gamma\right>_k$ vs. $V$ for fully connected graphs 
with $A=0$. As $V$ increases, the distribution is concentrated more around its mean 
value and an effective gap is created. The dashed line has slope $-0.25$ and is the best 
fit to the numerical data.}
\end{center}
\end{figure}

The resonance width distribution for a $V=5$ regular and generic graph is shown in
Fig.~\ref{fig2}b together with the RMT prediction, which reproduces the numerical 
distribution quite well. Fig.~\ref{fig3} shows a similar comparison for Neumann graphs. 
In this case, a clear deviation from the RMT prediction is observed. The main feature 
of ${\cal P}(\gamma)$ is the relatively high abundance of resonances in the vicinity 
of the real axis which is of dynamical nature and conforms with the expectations (see 
related discussion above Eq.~(\ref{gapKS})).

Although the absence of a gap in a strict sense is apparent in all the Neumann graphs
that we have studied, still one can speak about a gap in a ``probabilistic'' sense.
To show this, we have calculated the ratio of the resonance width standard deviation
$\Delta \gamma$ with respect to the mean resonance width $\left<\gamma\right>_k$ for
various fully connected graphs. Our numerical data reported in the inset of Fig.~\ref{fig3}
for various values of $V$ suggest that ${\Delta \gamma\over \left<\gamma\right>_k}
\sim {1\over V^{\alpha}}$ with $\alpha\simeq 0.25$.

For $g\gg 1$ (i.e. weak coupling regime) Eq.~(\ref{FSpole}) reduces to the following
expression
\begin{equation}
\label{chi2}
{\cal P}(\gamma) = \frac {(\beta M/2)^{\beta M/2}}{\langle \gamma\rangle_k{\it \Gamma}
(\beta M/2)} \left({\gamma\over\langle \gamma\rangle_k}\right)^{\beta M/2 -1}
\exp(-\gamma \beta M/2\langle \gamma\rangle_k ),
\end{equation}
where $\beta=1(2)$ for systems which respect (break) time-reversal symmetry, and
${\it \Gamma}(x)$ is the Gamma-function. We notice that (\ref{chi2}) is the so-called
$\chi^2$ distribution and can be derived independently using simple perturbation
theory \cite{JSA92}.

\begin{figure}
\begin{center}
\epsfxsize.5\textwidth%
\epsfbox{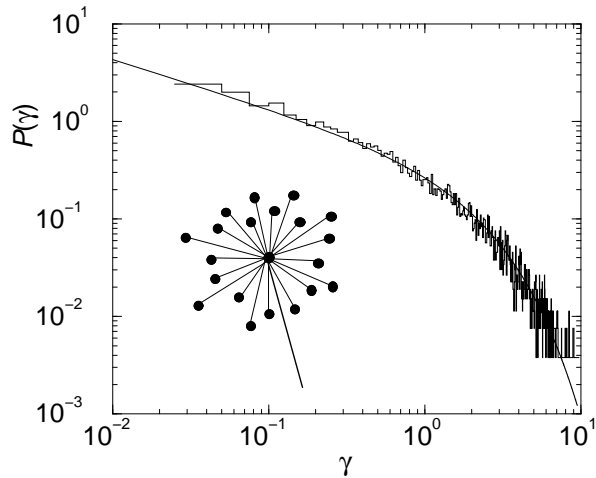}
\caption{\label{fig4} The rescaled resonance width distribution ${\cal P}(\gamma)$ for 
a star graph with $v_0=20$ (inset). The solid line is the RMT prediction Eq.~(\ref{chi2}).}
\end{center}
\end{figure}

A simple graph that satisfy the weak coupling limit is the Neumann star graph. They consist 
of $v_0$ bonds and one lead, all of which emanate from a single common vertex labeled with 
the index $i=0$. It is a simple matter to derive the scattering matrix $S^{(M=1)}$ which is 
reduced to a single phase factor \cite{KS00}
\begin{equation}
\label{contH}
S(k)\equiv {\rm e}^  {i\phi(k)} = {-\sum_{i=1}^{v_0} \tan(kL_i) +
i \over +\sum_{i=1}^{v_0} \tan(kL_i) + i}
\end{equation}

In Fig.~\ref{fig4} we present our numerical results for ${\cal P}(\gamma)$ for
a star graph with $v_0=20$. The data are in excellent agreement with the RMT
expectation given in Eq.~(\ref{contH}). We point out that in this case the
coupling to the continuum is weak since $g\simeq 10 \gg 1$ and therefore the
$\chi^2$-distribution with $M=1$ is applicable.

\subsection{\bf Resonance-trapping in graphs}

As mentioned at the end of Section \ref{rmat}, the effective Hamiltonian
Eq.~(\ref{heff}) is in general an infinite-dimensional and energy-dependent
matrix and thus very hard to treat. However, there are situations where this
Hamiltonian can be approximated with good accuracy by a finite and constant
matrix. For example, this is the case, when the energy levels of the closed
system are clustered, e.~g.\ by an underlying shell or band structure.

Let us calculate under this assumption the Ericson parameter Eq.~(\ref{eripar})
of a group of $N$
resonances of a graph at large $k\gg\Delta$. The resonances can be found by
diagonalizing $\cal H$. From Eq.~(\ref{heff}) we have for graphs with Neumann
b.~c.
\begin{eqnarray}\label{gammagraph}
\langle\gamma\rangle_{N}&=&
-{1\over kN\,\Delta}\mbox{Im}\,\mbox{tr}\,{\cal H}
\nonumber\\&=&
{\pi\over k N\Delta}\,\sum_{n,j}|W_{n,j}|^2
\nonumber\\&=&
M{\langle|\psi_{n,j}|^2\rangle\over \Delta}\,.
\end{eqnarray}
Note that diagonalization of $\cal H$ yields the resonances in the complex
energy plane, while we are interested in the wavenumber $\kappa$. This was
accounted for in the first line of Eq.~(\ref{gammagraph}). From the
normalization of the eigenfunctions of the graph we deduce that the mean
intensity of an eigenstate is the inverse of the total length ${\cal L}$ of the graph
$\langle|\psi_{n,j}|^2\rangle= {\cal L}^{-1}$. Hence, if $\Delta$ was equal to the
mean level spacing of all levels, $\pi/ {\cal L}$, the Ericson parameter would be
fixed to $\langle\gamma\rangle_N\sim M/\pi$.  However, we said above that the
effective Hamiltonian can be used only if there is clustering of levels,
meaning that the local mean spacing inside the cluster of $N$ resonances is
$\Delta\ll \pi/ {\cal L}$. Therefore we have in such a situation not only
$\langle\gamma\rangle_N\gg 1$ but even $\gamma\gg M$.

Theoretical calculations, based on RMT modeling, predict that in this
situation the statistical properties of the resonances change drastically:
with increasing strength of the coupling, $M$ very broad resonances are
formed, whereas the remaining $N-M$ approach the real axis and become long
lived (trapped) \cite{SZ88,HILSS92}. This counterintuitive phenomenon, which
was recently demonstrated in experiments on microwave resonators \cite{PRSB00}
is called ``resonance trapping''.

\begin{figure}
\begin{center}
\epsfxsize.5\textwidth%
\psfig{figure=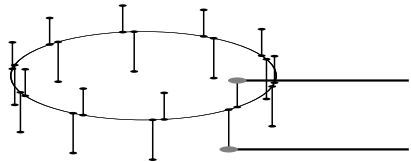,scale=0.5,angle=270}
\caption{\label{graph}Ring-shaped periodic graph with $N=10$ unit cells. }
\end{center}
\end{figure}

\begin{figure}
\begin{center}
\psfig{figure=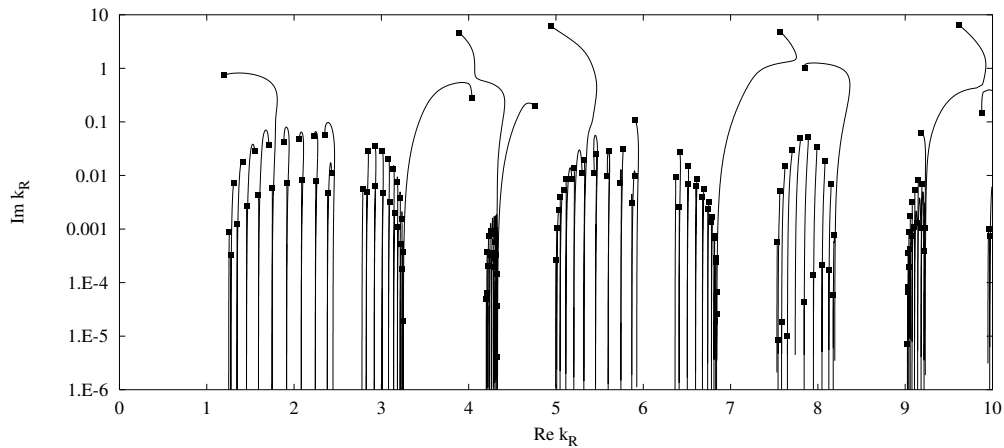,scale=0.7,angle=0}
\epsfxsize.5\textwidth%
\caption{\label{resotrap} Trajectories in the complex $\kappa -$plane of the resonances of a
ring graph with $N=20$ unit cells as $\lambda_{\rm leads}$ falls from $\infty$ to $0$. For 
the maximum coupling to the two attached scattering channels ($\lambda_{\rm leads} =0$), the 
location of the resonances is marked by squares. }
\end{center}
\end{figure}

Let us now illustrate this phenomenon for the case of quantum graphs. To this end we 
use a graph model consisting of $N$ identical unit cells forming a periodic ring as 
shown in Fig.~\ref{graph} (we note that our numerical data were already included in the 
recent review by Dittes \cite{Dit00}). Each unit cell consists of four vertices. Two 
of these vertices are $T$-shaped junctions with Neumann b.~c. ($\lambda=0$). The other 
two are dead ends where we impose Dirichlet b.~c. ($\lambda=\infty$). In one of the 
unit cells the dead ends have variable boundary conditions $0\le \lambda_{\rm leads} 
\le \infty$ and are connected to a scattering channel. As long as the coupling to the 
channels is zero ($\lambda=\infty$ at all dead ends), the spectrum is organized in 
discrete bands, each containing $N$ eigenvalues.  These eigenvalues are seen in 
Fig.~\ref{resotrap} as intersections of the solid lines with the real axis. We observe 
that the spacing within the bands can indeed be much smaller than the spacing between 
the bands.

As soon as $\lambda_{\rm leads}$ is finite, the eigenstates of the closed system are 
turned into resonances with a finite lifetime. In Fig.~\ref{resotrap} we show the motion 
of resonances in the complex $\kappa-$plane as we increase the coupling constant of 
the system by decreasing $\lambda_{\rm leads}$ to zero. One can clearly see the resulting
redistribution of the $S-$matrix poles. For particularly narrow bands, e.~g.\ at $k\sim 
4.2$, we observe indeed two broad resonances with a width several orders of magnitude 
above that of the remaining poles, while for the broader bands at least one broad 
resonance can be distinguished.

\section{\bf Conclusions}
\label{sec:conclusions}

Quantum graphs are used in our days a a paradigmatic model of quantum chaos.
By connecting them with infinite leads one turns them into scattering systems.
As a result, a scattering formalism can be developed, and an exact expression
for the scattering matrix can be found.

Due to the relative ease by which a large number of numerical data can be computed for
the graph models, we have performed a detailed statistical analysis of resonance
widths. A gap for the resonance widths has been obtained for ``generic" graphs and its 
absence was explained for Neumann graphs. Finally, we have shown that the phenomenon of
resonance trapping can appear also in quantum graphs.

\section{\bf Acknowledgments}

We would like to express our gratitude to Prof.~Uzy Smilansky to whom we owe
our interest in the subject and who contributed a great deal to most results
discussed in this paper. We also acknowledge many useful discussions with Y.
Fyodorov, A. Ossipov and M. Weiss. (T.~K) acknowledges support by a Grant from 
the GIF, the German-Israeli Foundation for Scientific Research and Development.


\end{document}